\documentclass[11pt]{article}

\usepackage{amsfonts,latexsym,eucal,color,graphicx,epsfig,amssymb,amsmath}

\newcommand{\be}{\begin{eqnarray}}
\newcommand{\ee}{\end{eqnarray}}
\newcommand{\pd}{\partial}
\newcommand{\nb}{\nabla}
\newcommand{\nn}{\nonumber}
\newcommand{\nl}{\nonumber \\ &&}

\newcommand{\dalm}{\kern1pt\vbox{\hrule height 0.9pt\hbox{\vrule width 0.9pt\hskip 2.5pt\vbox{\vskip 5.5pt}\hskip 3pt\vrule width 0.3pt}\hrule height 0.3pt}\kern1pt}
\newcommand{\ie}{{\it i.e.}}
\newcommand{\eg}{{\it e.g.}}
\newcommand{\etc}{{\it etc.}}
\newcommand{\et}{{\it et al.}}

\newcommand{\dd}{{\rm d}}

\makeatletter

\@addtoreset{equation}{section}
\makeatletter

\begin{document}

\begin{titlepage}

\begin{flushright}
{\small {\tt arXiv:1007.4302 [hep-th]}}\\
\end{flushright}
\vspace{1cm}
\begin{center}
{\Large  {\bf One-Dimensional Approximation of Viscous Flows}}\\
\vspace{.6cm}
Umpei Miyamoto\\
\vspace{.2cm}
{\it {\small Department of Physics, Rikkyo University, Tokyo 171-8501, Japan}}\\
\vspace{.2cm}
{\tt {\small umpei@rikkyo.ac.jp}}
\end{center}
\vspace{.2cm}
\begin{abstract}
Attention has been paid to the similarity and duality between the Gregory-Laflamme instability of black strings and the Rayleigh-Plateau instability of extended fluids. In this paper, we derive a set of simple (1+1)-dimensional equations from the Navier-Stokes equations describing thin flows of (non-relativistic and incompressible) viscous fluids. This formulation, a generalization of the theory of drop formation by Eggers and his collaborators, would make it possible to examine the final fate of Rayleigh-Plateau instability, its dimensional dependence, and possible self-similar behaviors before and after the drop formation, in the context of fluid/gravity correspondence.
\end{abstract}

\vspace{.5cm}

\tableofcontents

\end{titlepage}

\section{Introduction}
\label{sec:intro}

Black holes in higher dimensions have been studied from various points of view. In particular, they play practical and important roles in the contexts of large extra-dimension scenario and gauge/gravity correspondence. It has become known that they have rather richer structures than reported. Some explicit examples of instability and non-uniqueness in a solution space have been presented. In such a case, the phase transitions among various phases of black holes are expected (see \cite{Emparan:2008eg} for a review).

One important class of higher-dimensional black holes is that of Kaluza-Klein or `caged' black holes, \ie, black holes in a spacetime one of whose spatial dimensions is compactified on a circle, $\mathbf{S}^1$. Black objects known in this system include uniform black strings, non-uniform black strings, and black holes. Here, the black strings wrap the $\mathbf{S}^1$, whereas the black holes are localized in the $\mathbf{S}^1$-direction, and `uniform' stands for the translational invariance in the $\mathbf{S}^1$-direction. See \cite{Kol:2004ww,Harmark:2007md} for reviews. The uniform black strings were shown to be unstable by Gregory and Laflamme~\cite{Gregory:1993vy} with respect to linear perturbations breaking the translational invariance. Furthermore, the uniqueness is explicitly broken and the transitions among the three phases are expected. It is noted that the transition from a black string to a black hole via the pinch-off of horizon is accompanied by a topology change. Horowitz and Maeda~\cite{Horowitz:2001cz}, however, argued that the Gregory-Laflamme instability does not result in the pinch-off of horizon in a finite affine time. After this, a numerical simulation was performed by Choptuik \et~\cite{Choptuik:2003qd}, and many studies on the sequence of static solutions were done to figure out the phase diagrams.\footnote{Recently, Lehner and Pretorius reported the results of a new numerical simulation of a 5-dimensional black string~\cite{Lehner:2010pn}. They argue that the horizon indeed breaks up in a finite asymptotic time, accompanied by a self-similar cascade of the instability.}

The gauge/gravity or AdS/CFT (anti-de Sitter/conformal field theory) correspondence has provided the powerful tools to investigate strongly coupled quantum field theories by examining classical gravity. The behaviors of the field theories at finite temperature are encoded in the thermodynamics of black holes residing in higher dimensions. Thus, from the viewpoint of black hole physics, it is natural to expect that we could learn about the black holes from the field theory side. Regarding this point, there was important progress called the fluid/gravity correspondence. It was shown that the Einstein equations with a negative cosmological constant are dual to the Navier-Stokes equations for a certain kind of conformal fluid residing on the AdS boundary~\cite{Bhattacharyya:2008jc} (and see \cite{Rangamani:2009xk} for a review). This seems to open a new way to examine problems in black hole physics by working on the fluid mechanics which is a collective description of field theory at a long wavelength limit.

The similarities between the Gregory-Laflamme instability of black strings and the Rayleigh-Plateau instability of fluids supported by surface tension or Newtonian gravity were pointed out by Cardoso and his collaborators~\cite{Cardoso:2006ks,Cardoso:2006sj}, and then the similarities were found to persist indeed up to non-linear regimes~\cite{Miyamoto:2008rd}, at least as to the bifurcation structures in a phase diagram. One possibility of the physical origin of similarity is that the Gregory-Laflamme is holographically dual to the Rayleigh-Plateau instability of fluid lumps supported by surface tension, which is suggested by the conjecture that a lump of `gluon' plasma in large-$N$ gauge theories is dual to a black hole localized in an asymptotically AdS space with a Scherk-Schwarz compactification~\cite{Aharony:2005bm,Lahiri:2007ae}. It is natural to expect that in the theories allowing a first-order confine-deconfine phase transition the deconfined phase appears as localized balls separated from the confined phase by a certain kind of boundaries. The point of the above conjecture is that a gravitational dual of such a plasma ball is a finite-energy black hole (rather than an infinite-energy black brane) in the bulk (see Introduction of \cite{Cardoso:2009bv} for a brief but nice review). The phase diagrams of plasma tubes and plasma balls obtained in \cite{Maeda:2008kj} and \cite{Caldarelli:2008mv} with this approach mimic well those of the Kaluza-Klein black holes. These results suggest that we will be able to examine the dynamics of Kaluza-Klein black holes in non-linear regimes by working on the fluid mechanics, though the duality originally proposed in \cite{Aharony:2005bm,Lahiri:2007ae} is still a conjecture.\footnote{Another powerful and efficient way to study the Gregory-Laflamme instability is to solve the Einstein equations formulated in terms of an effective fluid that lives on a dynamical worldvolume, which is called the `blackfold' approach~\cite{Emparan:2009at}. The long-wavelength component of Gregory-Laflamme instability was obtained as a sound-mode instability~\cite{Emparan:2009at,Camps:2010br}. It is noted that this approach does not need a boundary of fluid, and so the instability is not Rayleigh-Plateau's one.}

The fluid equations involved in the fluid/gravity correspondence are the relativistic Navier-Stokes equations with specific values of transport coefficients (viscosities, thermal conductivity \etc) and an equation of state, that are much easier to treat than the dual gravitational equations. When the non-linear dynamics of a fluid is concerned, however, the fluid system is still difficult to treat as it is. Therefore, it is of practical importance to import and/or develop techniques in fluid mechanics to investigate the problems in black hole physics via the correspondence. In this paper, we import the scheme developed by Eggers and Dupont~\cite{EggersDupont} to describe the drop formation of free-surface flows (see \cite{Eggers:1997zz,EV} for comprehensive reviews), while generalizing it to that in arbitrary dimensions. The essence of this scheme is to reduce the ($2+1$)-dimensional partial differential equations and their boundary conditions describing axially symmetric flows to a set of ($1+1$)-dimensional equations by a radial-expansion method which is valid when the characteristic length scale in the direction perpendicular to the axis is smaller than that in the axial direction.\footnote{Needless to say, $N+1$ means $N$ spatial dimensions plus time. It is noted, however, that descriptions by ($N+1$)-dimensional partial differential equations is said $N$-dimensional (as the title of this paper) in non-relativistic physics.} It is noted that the approximate equations become exact as the pinch-off of a fluid neck is approached. In addition, we will see that with an appropriate prescription the approximate equations can describe relatively thick flows such as those at the onset of instability. By the vanishing of one independent variable, the problem is significantly simplified. Furthermore, in this approach the boundary conditions describing the balance among the surface tension, viscosity, and pressure jump at the fluid surface are contained in the set of reduced equations. In other words, the boundary conditions `disappear' from the system. While in this paper we will focus on a non-relativistic and incompressible limit for simplicity, the generalizations to relativistic and compressible cases would be straightforward (but see, \eg, \cite{Bhattacharyya:2008kq,Fouxon:2008tb,Eling:2009pb} for non-relativistic and incompressible fluids obtained via the fluid/gravity correspondence or the membrane paradigm).

In Sec.~\ref{sec:setup}, we introduce the Navier-Stokes equations and their boundary conditions, and the reduction to the axially symmetric flows is made. In Sec.~\ref{sec:1d}, the Navier-Stokes equations and their boundary conditions are combined to derive the set of ($1+1$)-dimensional equations, and their properties are investigated. In particular, the growth rate of the Rayleigh-Plateau instability of viscous flows are derived. Sec.~\ref{sec:conc} is devoted to discussions.

\section{Viscous flows with boundaries}
\label{sec:setup}

We review the Navier-Stokes equations and their boundary conditions (see, \eg, \cite{Landau:1987gn}), while generalizing them into those in arbitrary dimensions and curvilinear coordinates. See~\cite{Lahiri:2007ae,Caldarelli:2008mv,Cardoso:2009bv} for a relativistic formulation and especially \cite{Cardoso:2009bv} for a viscous case.

\subsection{Navier-Stokes equations}
\label{sec:NS}

We consider a fluid in a $d$-dimensional flat spacetime ($d \geq 4$) with a time coordinate $t$ and general curvilinear spatial coordinates $x^I$, ($I,J=1,2,\ldots,d-1$). The equation of continuity may be written as
\be
	\pd_t \rho + \nb_I ( \rho v^I ) = 0\ ,
\label{eq:EOC}
\ee
where $\rho$ is the mass density, $v^I$ the fluid velocity, $\nb_I$ a covariant derivative with respect to spatial metric $g_{IJ}(x^I)$.
The equations of motion for the fluid are obtained from the momentum conservation,
\be
	\pd_t ( \rho v^I ) + \nb_J \Pi^{IJ} = 0\ ,
\label{eq:mom-cons}
\ee
where $\Pi^{IJ}$ is a momentum flux density tensor.
The momentum flux density tensor in general may be written as
\be
	\Pi_{IJ}
	=
	\rho v_I v_J - \sigma_{IJ}\ ,
\ee
$\sigma_{IJ}$ being a stress tensor. We consider the viscous fluid whose stress tensor is given by
\be
	\sigma_{IJ}
	=
	- p g_{IJ}
	+
	\eta \left(
		\nb_I v_J + \nb_J v_I
		-
		\frac{2}{d-1} g_{IJ} \theta
	\right)
	+
	\zeta g_{IJ} \theta\ ,
\ee
where $p$ is the pressure; $\eta$ and $\zeta$ are the shear and bulk viscosities, respectively, assumed to be positive constants; $\theta := \nb_I v^I$ is the expansion.
Plugging the above expression of momentum flux density tensor into the conservation law (\ref{eq:mom-cons}), we obtain the Navier-Stokes equations with the help of equation of continuity~(\ref{eq:EOC}),
\be
	\rho
	D_t v_I
	=
	- \nb_I p + \eta \Delta v_I 
	+ \left(  \zeta + \frac{ d-3 }{ d-1 } \eta \right)
	\nb_I \theta\ ,
\label{eq:Navier-Stokes-1}
\ee
where the so-called substantial time derivative and Laplacian are defined by $D_t = \pd_t + v^I \nb_I  $ and $ \Delta = \nb^J \nb_J $, respectively.

Now, let us move to the boundary conditions imposed on the above equations when one considers two immiscible fluids (call them fluid 1 and fluid 2) divided by a surface. The position of the surface may be identified by the vanishing of a scalar function, $f(t,x^I) = 0$. Denoting the unit normal of the surface by $n_I = \nb_I f/ | \nb f | $ (pointing from fluid 1 to fluid 2) and surface tension by $\alpha(t,x^I)$, the stress balance equations, called the Young-Laplace formula, are given by
\be
	\left(
		\sigma^{(2)}_{IJ} - \sigma^{(1)}_{IJ}
	\right) n^J
	=
	\alpha \kappa n_I + P_{I}^{\;J} \nb_J \alpha
	\; \Big|_{f=0}\ ,
\label{eq:Balance1}
\ee
where $\kappa$ is [$( d-2 )$-times] the mean curvature of the surface and $P_{IJ}$ is the projection tensor, 
\be
	\kappa = \sum_{i=1}^{d-2} R_i^{-1} = \nb_I n^I\ ,
\;\;\;
	P_{IJ} = g_{IJ} - n_I n_J\ .
\ee 
Here, $R_i$ is the curvature radius (assumed to be positive if it is drawn in the side of fluid 1) in a principal direction on the surface. 

Here, we assume a few properties of the fluid and surface for simplicity. In the rest of this paper, we will focus on an incompressible fluid ($D_t \rho = 0$), for which  equation of continuity (\ref{eq:EOC}) and Navier-Stokes equations (\ref{eq:Navier-Stokes-1}) simplify,
\be
&&
	\nb_I v^I = 0\ ,
\label{eq:EOC2}
\\
&&
	D_t v_I
	=
	- \frac{ \nb_I p }{ \rho } + \nu \Delta v_I\ ,
\label{eq:Navier-Stokes-2}
\ee
$\nu=\eta/\rho$ being the kinetic viscosity.
Second, we assume that surface tension $\alpha$ is constant and neglect the effects of fluid 2 (\ie, $\sigma^{(2)}_{IJ} = 0$), in which case the normal and tangential components of stress-balance equation (\ref{eq:Balance1}) are given by (omitting the label of fluid 1)
\be
&&
	-  n_I \sigma^{IJ}  n_J
	=
	\alpha \kappa \; \big|_{f=0}\ ,
\label{eq:Balance2A}
\\
&&
	P_{I}^{\; J} \sigma_{JK} n^K = 0 \; \big|_{f=0}\ ,
\label{eq:Balance2B}
\ee 
where the stress tensor is reduced to
\be
	\sigma_{IJ}
	=
	- p g_{IJ} + \eta ( \nb_I v_J + \nb_J v_I )\ .
\ee

Finally, we present a kinetic boundary condition called the free-surface condition, stating that the surface moves with the fluid,
\be
	D_t f = 0 \; \big|_{f=0}\ .
\label{eq:KineticBC}
\ee
Hereafter, we will consider Eqs.~(\ref{eq:EOC2})-(\ref{eq:KineticBC}).

Under the above assumptions (\ie, the incompressibility, and the constancy of surface tension and viscosities), free parameters of the fluid are the density $ \rho $, kinetic viscosity $ \nu $, and surface tension $ \alpha $, which are dimensionally independent each other.
With these parameters, one can define typical length and time scales as\footnote{$\ell_\nu$ and $t_\nu$ may be called viscosity length and viscosity time, respectively. Their order of magnitude considerably depends on the kind of fluids (and interfaces), \eg, $\ell_\nu \sim 10^{-8} \; {\rm cm}$ for mercury and $\ell_\nu \sim 10^3 \; {\rm cm}$ for golden syrup~\cite{EV}.}
\be
	\ell_\nu := \frac{ \rho \nu^2 }{ \alpha }\ ,
\;\;\;
	t_\nu := \frac{ \rho^2 \nu^3 }{ \alpha^2 }\ .
\label{eq:vsc-length}
\ee
It is noted that one can regard, for example, ($ \rho, \ell_\nu, t_\nu $) as a set of three independent fluid constants rather than ($ \rho, \nu, \alpha $) by reversing (\ref{eq:vsc-length}),
\be
	\alpha = \frac{ \rho \ell_\nu^3 }{ t_\nu^2 }\ ,
\;\;\;
	\nu = \frac{ \ell_\nu^2 }{ t_\nu }\ .
\label{eq:alphanu}
\ee

\subsection{Axially symmetric flows}
\label{sec:axial}

\begin{figure}[t!]
	\begin{center}
			\includegraphics[width=8cm]{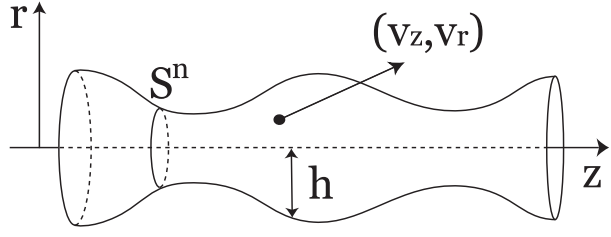}
	\caption{{\sf A schematic figure of the axially symmetric flow in $(n+2)$-dimensional space.}}
	\label{fg:flow}
	\end{center}
\end{figure}

Since we are concerned with the dynamics of axially symmetric flows, it is convenient to introduce a cylindrical coordinate system with which the line element of ($d-1$)-dimensional flat space is written as
 \be
	g_{IJ} (x^I) \dd x^I \dd x^J
	=
	\dd z^2 + \dd r^2 + r^2 \dd \Omega_n^2
	=
	\delta_{ab} \dd x^a \dd x^b+ r^2 \gamma_{ij}(\phi^i) \dd \phi^i \dd \phi^j\ ,
\ee
where $ \dd \Omega_n^2 = \gamma_{ij}(\phi) \dd \phi^i \dd \phi^j $  ($ i,j = 1,2,\ldots,n:=d-3$) is the line element of a unit $n$-sphere and $ x^a = (z,r) $.
Now, let us assume that the fluid has a velocity only in the directions of $z$ and $r$, and that the fluid surface respects the axial symmetry. That is, the velocity field $v^I(t,x^I)$ and the scalar function $f(t,x^I)$ are written as
\be
	v^a = v^a (t,z,r)\ ,
\;\;\;
	v^i = 0\ ,
\;\;\
	f = r - h(t,z)\ ,
\ee
where $h(t,z)$ is called a height function, giving the local radius of flow (see Fig.~\ref{fg:flow}).

Then, equation of continuity (\ref{eq:EOC2}) reduces to
\be
	\pd_z v_z + \left( \pd_r + \frac{n}{r} \right) v_r = 0\ .
\label{eq:eoc}
\ee
$z$ and $r$ components of Navier-Stokes equation (\ref{eq:Navier-Stokes-2})
are given by\footnote{Non-zero components of Levi-Civita connection $\Gamma^{I}_{JK}$ are
\be
	\Gamma^a_{ij} = -r \delta^a_r \gamma_{ij}\ ,
\;\;\;\;
	\Gamma^i_{aj} = r^{-1} \delta_a^r \delta^i_j\ ,
\;\;\;\;
	\Gamma^{i}_{jk} = {}^{(\gamma)}\Gamma^i_{jk}\ ,
\ee
where ${}^{(\gamma)}\Gamma^i_{jk}$ is the connection with respect to the metric of unit $n$-sphere. With using these connections, the vector derivatives in the Navier-Stokes equation are calculated. For example,  
\be
	\nb_I v_J
	=
	\pd_a v_b \delta^a_I \delta^b_J 
	+
	r v_r \gamma_{ij} \delta^i_I \delta^j_J\ ,
\;\;\;
	\Delta v_a
	=
	( \pd_z^2 + \pd_r^2 ) v_a
	+ \frac{n}{r} \pd_r v_a 
	- \frac{n}{r^2} \delta_a^r v_r\ .
\ee
}
\be
	\pd_t v_z + ( v_z \pd_z + v_r \pd_r ) v_z
	&=&
	- \frac{ \pd_z p }{ \rho }
	+ \nu
	\left(
		\pd_z^2 + \pd_r^2 + \frac{n}{r} \pd_r 
	\right) v_z\ ,
\label{eq:ns1}
\\
	\pd_t v_r + ( v_z \pd_z + v_r \pd_r ) v_r
	&=&
	- \frac{ \pd_r p }{ \rho }
	+ \nu
	\left(
		\pd_z^2 + \pd_r^2
		+
		\frac{n}{r} \pd_r - \frac{n}{r^2} 
	\right) v_r\ .
\label{eq:ns2}
\ee
Normal and tangential stress-balance equations (\ref{eq:Balance2A}) and (\ref{eq:Balance2B}) reduce to
\be
	\frac{p}{\rho}
	-
	\frac{2\nu}{1 + ( \pd_z h )^2}
	\Big[
		( \pd_z h )^2 \pd_z v_z 
		-
		(\pd_z h) ( \pd_z v_r + \pd_r v_z )
		+
		\pd_r v_r
	\Big]
	&=&
	\frac{ \alpha }{ \rho } \kappa \; \Big|_{r=h}\ ,
\label{eq:sb1}
\\
	\frac{\nu}{ 1+ (\pd_z h)^2 }
	\Big[
		2  (\pd_z h) ( \pd_z v_z  -  \pd_r v_r )
		- \left[ 1- ( \pd_z h )^2 \right]  ( \pd_z v_r + \pd_r v_z   )
	\Big]
	&=&
	0 \; \Big|_{r=h}\ ,
\label{eq:sb2}
\ee
where the mean curvature is given by
\be
	\kappa
	=
	- \frac{ \pd_z^2 h }{ \left[ 1+ (\pd_z h)^2 \right]^{3/2} }
	+ \frac{ n }{ r \sqrt{ 1+(\pd_z h)^2 } }\ .
\label{eq:kappa}
\ee
It is noted that the $z$ and $r$ components of tangential stress-balance equation (\ref{eq:Balance2B}) lead to the same result, Eq.~(\ref{eq:sb2}), due to the isotropy of surface tension.
Finally, kinetic boundary condition (\ref{eq:KineticBC}) reduces to
\be
	- \pd_t h - v_z \pd_z h + v_r = 0 \; \Big|_{r=h}\ .
\label{eq:kbc}
\ee

\section{One-dimensional approximation}
\label{sec:1d}

In this section, we derive a set of ($1+1$)-dimensional equations from the ($2+1$)-dimensional equations obtained in Sec.~\ref{sec:axial}, while generalizing the formulation in \cite{EggersDupont} (details are given in \cite{Eggers:1997zz}).

\subsection{Radial expansion}
\label{sec:expansion}

Let us consider the situation in which a characteristic length scale in the $r$-direction (typically, the thickness of the fluid neck), denoted by $\ell_r$, is smaller enough than a characteristic length scale in the axial direction, denoted by $\ell_z$. We expect in this situation that the dynamics is dominated by the motion in the $z$-direction. Assuming this situation, we define a small dimensionless parameter $\epsilon$ and a typical time scale $\tau$ determined by the motion in the $z$-direction,
\be
	\epsilon
	\sim
	\frac{ \ell_r }{ \ell_z }\ ,
\;\;\;
	\tau
	\sim
	\frac{ \ell_z }{ v_z }\ .
\label{eq:EpsilonTau}
\ee
As will be seen below, we can show that the length and time scales, $\ell_z$ and $\tau$, in such a situation have the following order of magnitudes measured in the fluid constants (\ref{eq:vsc-length}),
\be
	\ell_z \sim \epsilon \ell_\nu\ ,
\;\;\;
	\tau \sim \epsilon^2 t_\nu\ ,
\label{eq:order}
\ee
if one demands the balance among the contributions to the dynamics from the inertia, pressure gradient, viscosity, and surface tension. 

First, we demand the balance among the three contributions to Navier-Stokes equation (\ref{eq:ns1}) from the inertia, pressure gradient, and viscosity:
\be
	\pd_t v_z
	\sim
	\frac{ \pd_z p }{ \rho }
	\sim
	\nu \pd_z^2 v_z\ .
\label{eq:NS-balance}
\ee
Replacing the derivatives in Eq.~(\ref{eq:NS-balance}) by corresponding scales, we have two relations for $\ell_z$ and $\tau$,
\be
	\frac{ \ell_z^2 }{ \tau^2 }
	\sim
	\frac{ p }{ \rho }\ ,
\;\;\;
	\frac{ \ell_z^2 }{ \tau }
	\sim
	\nu\ .
\label{eq:ellzTau}
\ee
One can estimate the order of pressure by the equation of normal-stress balance (\ref{eq:sb1}), namely $ p \sim \alpha \kappa $. In addition, the mean curvature of a sufficiently flat surface is dominated by the curvature in the spherical part [\ie, the second term on the right-hand side of Eq.~(\ref{eq:kappa})] to yield $ \kappa \sim 1/\ell_r \sim 1/( \epsilon \ell_z )$. Thus, we have
\be
	p
	\sim 
	\frac{ \alpha }{  \epsilon \ell_z }\ .
\label{eq:p-tension}
\ee
Eliminating the pressure in (\ref{eq:ellzTau}) by (\ref{eq:p-tension}), one can immediately solve (\ref{eq:ellzTau}) for $\ell_z$ and $\tau$, to obtain Eq.~(\ref{eq:order}).

From the above argument, we can estimate the order of magnitudes of all quantities in our basic equations in Sec.~\ref{sec:axial}. Namely, the magnitude of each quantity can be expressed in terms of dimensionless parameter $\epsilon$ and fluid constants ($ \rho, \ell_\nu, t_\nu $). With this in mind, we introduce the following dimensionless variables with hats,
\be
&&
	t = \epsilon^2 t_\nu \hat{t}\ ,
\;\;\;
	z = \epsilon \ell_\nu \hat{z}\ ,
\;\;\;
	r = \epsilon^2 \ell_\nu \hat{r}\ ,
\;\;\;
	h 
	=
	\epsilon^2 \ell_\nu \hat{h}\ ,
\nl
	v_z
	=
	\frac{ \ell_\nu }{ \epsilon t_\nu } \hat{v}_z\ ,
\;\;\;
	v_r
	=
	\frac{ \ell_\nu }{ t_\nu } \hat{v}_r\ ,
\;\;\;
	p
	=
	\frac{  \rho \ell_\nu^2 }{ \epsilon^2 t_\nu^2  }  \hat{p}\ ,
\;\;\;
	\kappa
	=
	\frac{ 1 }{ \epsilon^2 \ell_\nu } \hat{\kappa}\ .
\label{eq:hats}
\ee

Plugging Eqs.~(\ref{eq:alphanu}) and (\ref{eq:hats}) into the basic equations, all equations are made dimensionless. Equation of continuity (\ref{eq:eoc}) does not change its form,
\be
	\hat{\pd}_{z} \hat{v}_z
	+
	\left(
		\hat{\pd}_{ r }
		+
		\frac{ n }{ \hat{r} }
	\right) \hat{v}_r
	=
	0\ .
\label{eq:dl-eoc}
\ee
Navier-Stokes equations (\ref{eq:ns1}) and (\ref{eq:ns2}) lead to
\be
	\hat{\pd}_t \hat{v}_z 
	+ 
	( \hat{v}_z \hat{\pd}_z + \hat{v}_r \hat{\pd}_r ) \hat{v}_z
	&=&
	- \hat{\pd}_z \hat{p}
	+ 
	\left[
		\hat{\pd}_z^2
		+ \frac{1}{\epsilon^2}
		\left(
			\hat{\pd}_r^2 + \frac{n}{ \hat{r} } \hat{\pd}_r
		\right)
	\right] \hat{v}_z\ ,
\label{eq:dl-ns1}
\\
	\hat{\pd}_t \hat{v}_r
	+
	( \hat{v}_z \hat{\pd}_z + \hat{v}_r \hat{\pd}_r ) \hat{v}_r
	&=&
	- \frac{ 1 }{ \epsilon^2 } \hat{\pd}_r \hat{p}
	+
	\left[
		\hat{\pd}_z^2
		+ \frac{ 1 }{ \epsilon^2 }
		\left(
			\hat{\pd}_r^2
			+ \frac{n}{ \hat{r} } \hat{\pd}_r
			- \frac{n}{ \hat{r}^2 }
		\right)
	\right] \hat{v}_r\ .
\label{eq:dl-ns2}
\ee
Normal and tangential stress-balance equations (\ref{eq:sb1}) and (\ref{eq:sb2}) become
\be
	\hat{p}
	-
	\frac{2}{1 + \epsilon^2 ( \hat{\pd}_z \hat{h} )^2}
	\left[
		\epsilon^2 ( \hat{\pd}_z \hat{h} )^2 \hat{\pd}_z \hat{v}_z 
		-
		( \hat{\pd}_z \hat{h} )
		\left( \epsilon^2 \hat{\pd}_z \hat{v}_r + \hat{\pd}_r \hat{v}_z \right)
		+
		\hat{\pd}_r \hat{v}_r
	\right]
	&=&
	\hat{\kappa} \; \Big|_{\hat{r}=\hat{h}}\ ,
\label{eq:dl-sb1}
\\
		\frac{1}{ 1+ \epsilon^2 ( \hat{\pd}_z \hat{h} )^2 }
	\left[
		2  \epsilon^2 (\hat{\pd}_z \hat{h}) ( \hat{\pd}_z \hat{v}_z  -  \hat{\pd}_r \hat{v}_r )
		- \left[ 1- \epsilon^2 ( \hat{\pd}_z \hat{h} )^2 \right]  ( \epsilon^2 \hat{\pd}_z \hat{v}_r + \hat{\pd}_r \hat{v}_z   )
	\right]
	&=&
	0 \; \Big|_{\hat{r}=\hat{h}}\ ,
\label{eq:dl-sb2}
\ee
where the mean curvature is given by
\be
	\hat{\kappa}
	=
	- \frac{ \epsilon^2 \hat{\pd}_z^2 \hat{h} }
		   { \left[ 1+ \epsilon^2 (\hat{\pd}_z \hat{h})^2 \right]^{3/2} }
	+ \frac{ n }{ \hat{r} \sqrt{ 1+  \epsilon^2 (\hat{\pd}_z \hat{h})^2 } }\ .
\label{eq:dl-kappa}
\ee
Finally, kinetic boundary condition (\ref{eq:KineticBC}) does not change its form
\be
	- \hat{\pd}_t \hat{h}
	- \hat{v}_z \hat{\pd}_z \hat{h}
	+ \hat{v}_r
	=
	0 \; \Big|_{\hat{r}=\hat{h}}\ .
\label{eq:dl-kbc}
\ee

Now, we are ready to derive an effective $(1+1)$-dimensional equations by approximating the dependence of fluid quantities on $\hat{r}$ by Taylor-series expansion.
First, the velocity in the $z$-direction and the pressure are expanded as
\be
	\hat{v}_z (\hat{t},\hat{z},\hat{r})
	&=&
	\hat{v}_0(\hat{t},\hat{z})
	+ \hat{v}_2(\hat{t},\hat{z}) ( \epsilon \hat{r} )^2
	+  O\left( ( \epsilon \hat{r} )^4 \right)\ ,
\nn
\\
	\hat{p} (\hat{t},\hat{z},\hat{r})
	&=&
	\hat{p}_0(\hat{t},\hat{z})
	+ \hat{p}_2(\hat{t},\hat{z}) ( \epsilon \hat{r} )^2
	+ O\left( ( \epsilon \hat{r} )^4 \right)\ .
\label{eq:expansion}
\ee
Then, the velocity in the $r$-direction is determined order-by-order from the equation of continuity (\ref{eq:dl-eoc}),
\be
	\epsilon \hat{v}_r (\hat{t},\hat{z},\hat{r})
	=
	- \frac{1}{n+1} \hat{\pd}_{z} \hat{v}_0 (\epsilon \hat{r}) 
	- \frac{1}{n+3} \hat{\pd}_{z} \hat{v}_2 ( \epsilon \hat{r})^3
	+ O\left( ( \epsilon \hat{r} )^5 \right)\ .
\ee
The leading order of Navier-Stokes equation in the $r$-direction (\ref{eq:dl-ns2}) gives
\be
	\hat{\pd}_{t} \hat{v}_0 + \hat{v}_0 \hat{\pd}_{z} \hat{v}_0
	=
	- \hat{\pd}_{z} \hat{p}_0
	+ 2(n+1) \hat{v}_2 + \hat{\pd}_{z}^2 \hat{v}_0\ .
\label{eq:v0-eq-tmp}
\ee
The leading orders of normal and tangential stress-balance equations, (\ref{eq:dl-sb1}) and (\ref{eq:dl-sb2}), give
\be
&&
	\hat{p}_0
	+
	\frac{2}{n+1} \hat{\pd}_z \hat{v}_0
	=
	\hat{ \kappa }\ ,
\label{eq:p0-eq}
\\
&&
	2(n+2) \hat{\pd}_z \hat{h} \hat{\pd}_z \hat{v}_0 
	+ \hat{h} \hat{\pd}_z^2 \hat{v}_0
	- 2 (n+1)  \hat{h} \hat{v}_2 = 0\ ,
\label{eq:v2-eq}
\ee
where the leading order of dimensionless mean curvature is given by
\be
	\hat{\kappa} = \frac{ n }{ \hat{h} }\ .
\label{eq:lowest-kappa}
\ee
Eliminating $\hat{p}_0$ and $\hat{v}_2$ from Eq.~(\ref{eq:v0-eq-tmp}) with using Eqs.~(\ref{eq:p0-eq}) and (\ref{eq:v2-eq}), we obtain an equation of motion for $\hat{v}_0$,
\be
	\hat{\pd}_t \hat{v}_0 + \hat{v}_0 \hat{\pd}_z \hat{v}_0
	=
	- \hat{\pd}_z \hat{\kappa}
	+ \frac{ 2(n+2) }{ n+1 } 
	\frac{ \hat{\pd}_z (  \hat{h}^{n+1} \hat{\pd}_z \hat{v}_0 ) }
		 { \hat{h}^{n+1} }\ .
\label{eq:v-eq}
\ee
Finally, from the leading order of kinetic boundary condition (\ref{eq:dl-kbc}), we obtain an equation of motion for $\hat{h}$,
\be
	\hat{\pd}_t \hat{h} + \hat{v}_0 \hat{\pd}_z \hat{h}
	+
	\frac{ 1 }{ n+1 } \hat{h} \hat{\pd}_z \hat{v}_0 
	=
	0\ .
\label{eq:h-eq}
\ee

We have obtained a closed set of ($1+1$)-dimensional equations, (\ref{eq:v-eq}) and (\ref{eq:h-eq}), that describes the flows of viscous fluid in general dimensions.
Apart from the simplification that independent variable $r$ is eliminated from the system, the leading-order effect of the stress-balance equations at the surface is encoded in the first term of the right-hand side of Eq.~(\ref{eq:v-eq}) and the stress-balance equations themselves apparently disappear from the system.

\subsection{Conservation laws}
\label{sec:cons}

Let us write Eqs.~(\ref{eq:v-eq}) and (\ref{eq:h-eq}) in dimensionful form. Writing the corresponding dimensionful quantity without a hat and putting $v(t,z) := v_0(t,z)$, we obtain a set of equations for $v(t,z)$ and $ h(t,z) $,
\be
&&
	\dot{v} + v v^\prime 
	=
	- \frac{ \alpha \kappa^\prime }{ \rho }
	+ \frac{ 2(n+2) \nu }{ n+1 }
	  \frac{ ( h^{n+1} v^\prime )^\prime }{ h^{n+1} }\ ,
\label{eq:master1}
\\
&&
	\dot{h} + v h^\prime + \frac{1}{n+1} h v^\prime = 0\ ,
\label{eq:master2}
\ee
where the dot and prime denote the derivatives with respect to $t$ and $z$, respectively.
From Eq.~(\ref{eq:lowest-kappa}), the mean curvature is given by
\be
	\kappa
	= \frac{n}{h}\ .
\label{eq:kappa3}
\ee

Although the greatest significance of the system (\ref{eq:master1}) and (\ref{eq:master2}) lies in the asymptotic limit of $\epsilon \to 0$, it has been successfully used for simulations {\it away} from the breakup (e.g., in \cite{EggersDupont}). The key to this success lies in a modification of Eq.~(\ref{eq:kappa3}) to the complete expression of mean curvature [see Eq.~(\ref{eq:kappa})],
\be
	\kappa
	=
	- \frac{ h^{\prime \prime} }
		 { \left(
				1 + h^{\prime 2}
		   \right)^{3/2}  }
	+ \frac{ n }{ h \sqrt{ 1 + h^{\prime 2} } }\ .
\label{eq:kappa2}
\ee
The right-hand side of Eq.~(\ref{eq:kappa2}) includes an infinite sequence of terms of the radial expansion. The above prescription has not been justified completely so far and is `phenomenological' at this point. It is highly desirable to develop a more general and consistent method to `resum' all the relevant terms of the expansion. Readers interested in this point are directed to Sec.~V.B of \cite{Eggers:1997zz}, where additional rationale for this prescription is discussed. Anyway, we will investigate the dynamical properties of the system described by Eqs.~(\ref{eq:master1}) and (\ref{eq:master2}) with the {\it complete} expression of mean curvature, Eq.~(\ref{eq:kappa2}). We will see that the above replacement indeed broadens the applicability of Eqs.~(\ref{eq:master1}) and (\ref{eq:master2}).

In order to see the structure of Eqs.~(\ref{eq:master1}) and (\ref{eq:master2}),
it is instructive to rewrite them as follow,
\be
&&
	\pd_t ( h^{n+1} v ) + \pd_z ( h^{n+1} v^2 )
	=
	- \frac{\alpha \kappa^\prime}{\rho}  h^{n+1}
	+ \frac{2(n+2)}{n+1} \nu ( h^{n+1} v^\prime )^\prime\ ,
\label{eq:master1'}
\\
&&
	\pd_t ( h^{n+1} ) + \pd_z ( h^{n+1} v ) = 0\ .
\label{eq:master2'}
\ee
The form of Eq.~(\ref{eq:master1'}) is best motivated by considering the force balance on a slice of fluid which carries linear momentum per unit length $ h^{n+1} v $, disregarding the common factor $\rho \Omega_n$, where $\Omega_n := 2\pi^{(n+1)/2} / \Gamma [ (n+1)/2 ] $ is the area of unit $n$-sphere.
The left-hand side is the total time derivative of the momentum. The first term on the right-hand side comes from the capillary forcing on the slice, the second term is the viscosity forcing. The meaning of Eq.~(\ref{eq:master2'}) is also clear: it is written as a conservation law for the mass per unit length $h^{n+1}$, disregarding the common factor $\rho \Omega_n$ again.

Now, we express the mass conservation and energy dissipation due to the viscosity in integral form. We write the mass in a fixed interval $ z \in [z_-,z_+] $ as
\be
	M(t)
	=
	\rho \Omega_{n}
	\int_{z_-}^{z_+} h^{n+1} dz\ ,
\ee
Then, one can show that from Eq.~(\ref{eq:master2}) this mass satisfies a conservation law,
\be
	\pd_t  M
	=
	- \rho \Omega_{n}
	\left(
		h^{n+1} v
	\right) \Big|_{z_-}^{z_+}\ .
\label{eq:mass-cons}
\ee
The kinetic energy and the potential energy stored in the surface may be given by
\be
	E_{\rm{kin}}(t)
	=
	\frac{1}{2} \rho \Omega_n
	\int_{z_-}^{z_+} h^{n+1} v^2 dz\ ,
\;\;\;
	E_{\rm{surf}}(t)
	=
	(n+1) \Omega_n \alpha 
	\int_{z_-}^{z_+} h^n \sqrt{ 1+h^{\prime 2} } dz\ .
\ee
Combining Eqs.~(\ref{eq:master1}), (\ref{eq:master2}), and (\ref{eq:kappa2}), one can show
\be
&&
	\pd_t ( E_{\rm{kin}} + E_{\rm{surf}} )
	=
	- \frac{ 2(n+2)\Omega_n }{ n+1 }
	\rho \nu 
	\int_{z_-}^{z_+} h^{n+1} v^{\prime 2} dz
\nn
\\
&& \hspace{1cm}
	-
	\Omega_n
	\left(
		\frac12 \rho h^{n+1} v^3
		-
		\frac{2(n+2)}{n+1} \rho \nu h^{n+1}  v v^\prime
		+
		\alpha h^{n+1} v \kappa
		-
		(n+1) \alpha \frac{ h^n h^\prime \dot{h} }{ \sqrt{ 1+ h^{\prime 2} } }  
	\right) \Bigg|_{z_-}^{z_+}\ .
\label{eq:energy-diss}
\ee
This means that apart from the driving through the boundary terms, the total energy of the system can only decrease due to the viscosity, and the fluid eventually has to reach some static equilibrium shape corresponding to a minimum of the surface energy. It is noted that Eq.~(\ref{eq:kappa2}) is essential to derive Eq.~(\ref{eq:energy-diss})\footnote{We have used an identity,
\be
	n h^{n-1} \sqrt{ 1+h^{\prime 2} }
	=
	h^n \kappa
	+
	\left(
		\frac{ h^n h^\prime }{ \sqrt{ 1+h^{\prime 2} } }
	\right)^\prime \ ,
\ee
which holds for Eq.~(\ref{eq:kappa2}) but not for Eq.~(\ref{eq:kappa3}).} and to have the above physically reasonable picture on the final equilibrium, which will be realized far from the asymptotic limit of $\epsilon \to 0$.

\subsection{Instability of viscous flows}
\label{sec:RP}

The dispersion relation or growth rate of Rayleigh-Plateau instability of viscous flows can be derived analytically with using the above formulation.

Let us consider linear perturbations of a static cylindrical tube whose radius is $r_0$,
\be
	h (t,z) = r_0 \left[ 1 + \delta e^{\omega t} \cos ( kz ) \right]\ ,
\;\;\;
	v (t,z) = \delta V_0  e^{\omega t} \sin(kz)\ ,
\ee
where $\delta$ and $V_0$ are constants ($|\delta| \ll 1$). Plugging these expansions into Eqs.~(\ref{eq:master1}) and (\ref{eq:master2}), we obtain algebraic equations for $\omega$ and $V_0$ at $O(\delta)$, which are solved to yield
\be
&&
	\omega(k)
	=
	\frac{n+2}{n+1} \omega_0
	\left(
		\sqrt{
			\frac{n+1}{(n+2)^2} (kr_0)^2 \left[ n-(kr_0)^2 \right]
			+
			\frac{\ell_\nu}{r_0} ( kr_0 )^4
		}
		-
		\sqrt{ \frac{\ell_\nu}{r_0} } ( kr_0 )^2
	\right)\ ,
\label{eq:disp}
\\
&&
	V_0
	=
	-(n+2)
	\frac{ \ell_\nu^2 }{ t_\nu r_0 }
	\left(
		\sqrt{
			\frac{n+1}{(n+2)^2} \frac{ r_0 }{ \ell_\nu }
			[ n-(kr_0)^2 ]
			+
			( kr_0 )^2
		}
		-kr_0
	\right)\ ,
\;\;\;
	\omega_0^2 = \frac{ \alpha }{ \rho r_0^3 }\ .
\ee
The growth rate in the low-viscosity ($\ell_\nu/r_0 \ll 1$) and high-viscosity ($\ell_\nu/r_0 \gg 1$) limits are given by
\be
	\omega
	\simeq
	\frac{ \omega_0 }{ \sqrt{ n+1 } } ( kr_0 ) \sqrt{ n-(kr_0)^2 }
\;\;\;\;\;
\mbox{and}
\;\;\;\;\;
	\omega
	\simeq
	\frac{ \omega_0 }{ 2(n+2) }
	\left( \frac{r_0}{\ell_\nu} \right)^{1/2}
	\left[ n-(kr_0)^2 \right]\ ,
\label{eq:hilow-limits}
\ee
respectively.
Both the above limits for $n=1$ coincide with the classic result of Chandrasekhar~\cite{Chandrasekhar} if an expansion to lowest order in $kr_0$ is made.

\begin{figure}[t!]
	\begin{center}
		\setlength{\tabcolsep}{ 5 pt }
		\begin{tabular}{ cc }
			{\sf (a)} & {\sf (b)} \\
			\includegraphics[width=8cm]{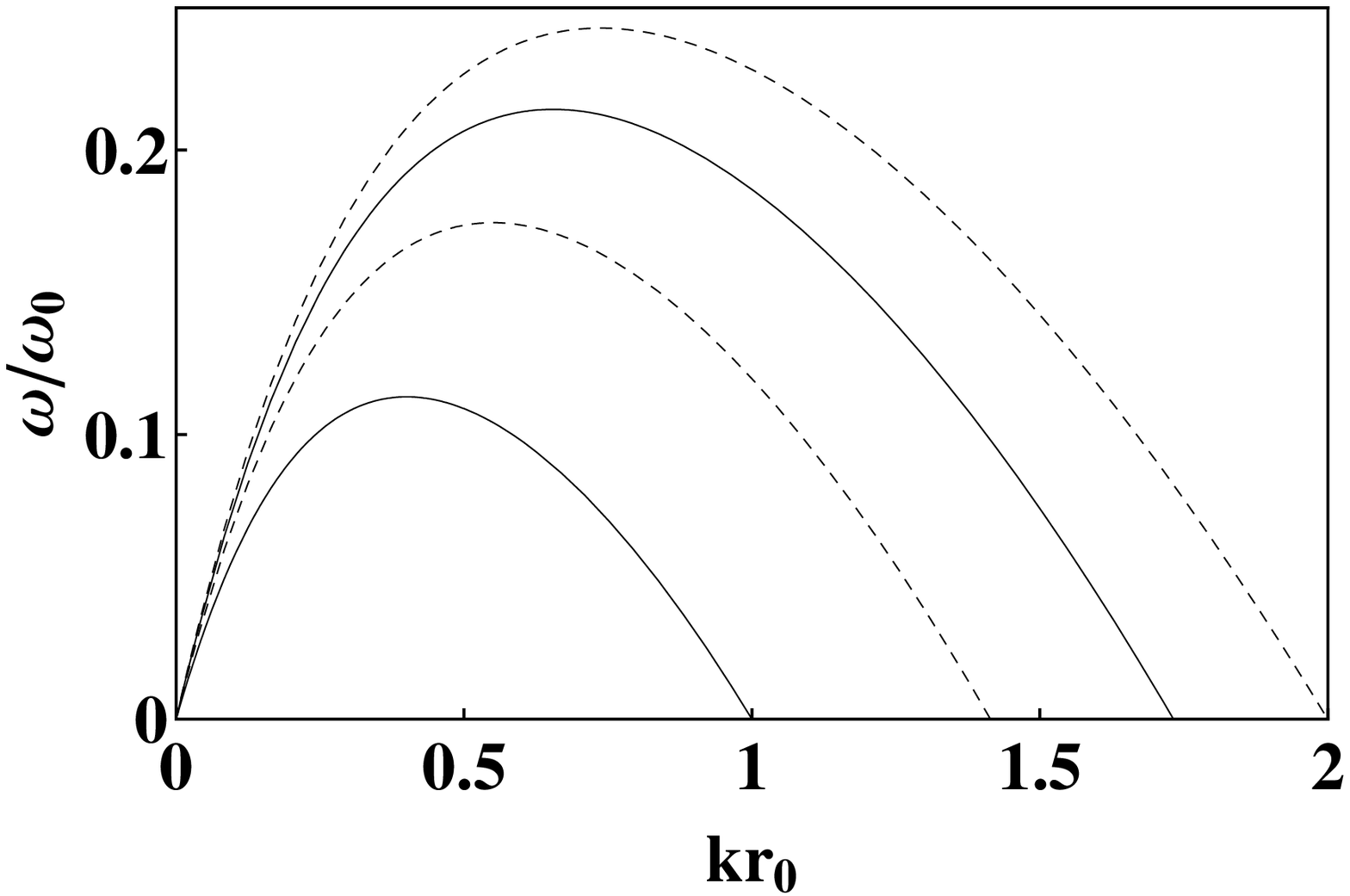} &
			\includegraphics[width=8cm]{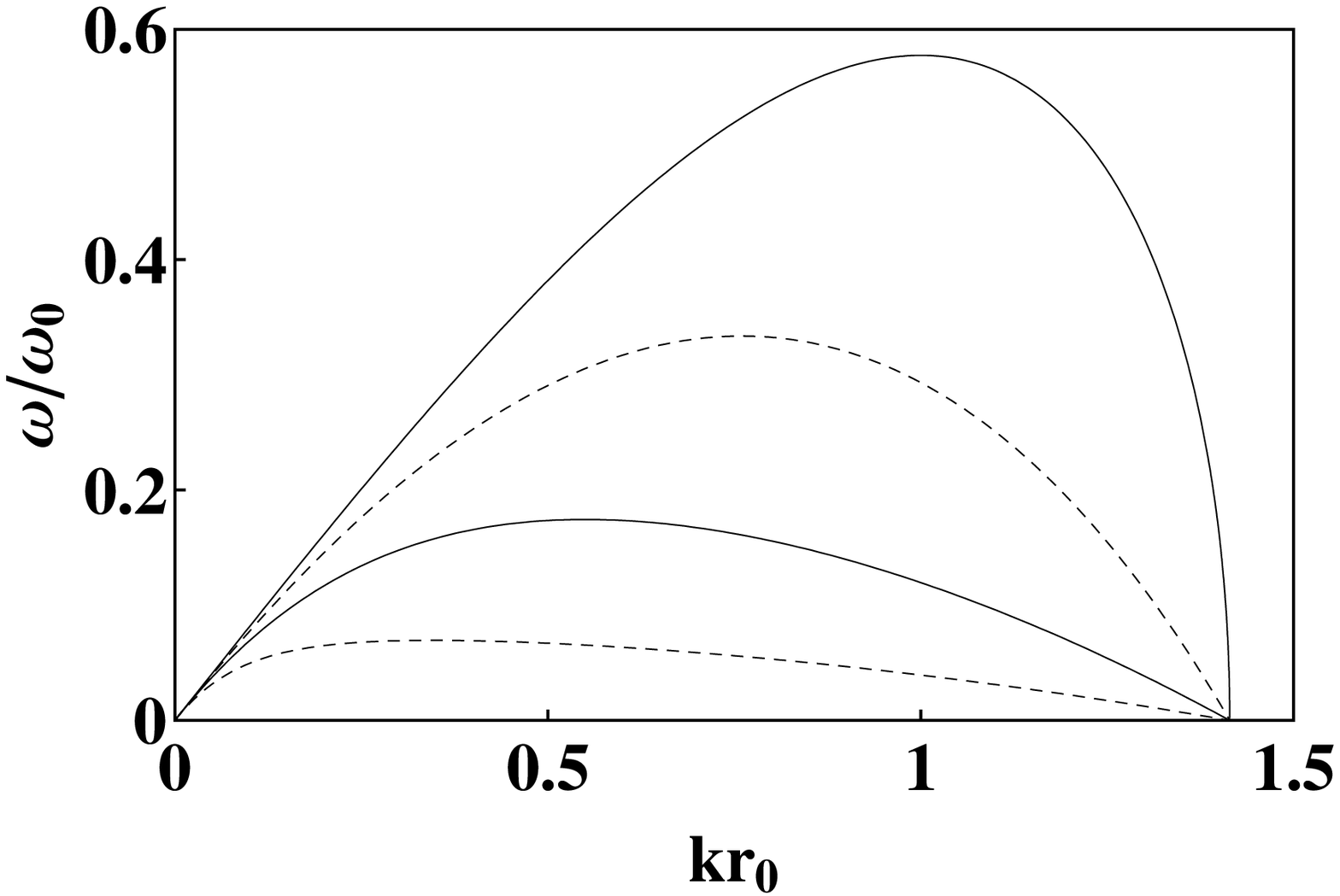} \\
		\end{tabular}
	\caption{{\sf (a) The dimensional dependence of the growth rate of Rayleigh-Plateau instability for $\ell_\nu / r_0 = 1$. $d=4,5,6,7$ from the bottom to the top. (b) The viscosity dependence for $d=5$. $\ell_\nu / r_0 = 0, 0.1, 1, 10$ from the top to the bottom.}}
	\label{fg:disp}
	\end{center}
\end{figure}

One sees from Eq.~(\ref{eq:disp}) that the wavenumber of the onset of instability [$k_c>0$, $ \omega(k_c) = 0 $] is given by $ k_c = \sqrt{n}/r_0 $, irrespective of the viscosity, which generalizes the result of the inviscid case~\cite{Cardoso:2006ks} and is similar to the dimensional dependence of Gregory-Laflamme instability~\cite{Kol:2004pn,Asnin:2007rw}. One the other hand, the wavenumber of the most unstable or fastest growing mode $k_{\rm max}$ is a decreasing function of viscosity,
\be
	( k_{\rm{max}} r_0)^2 
	=
	\frac{ n }{ 2 \left[  1 + (n+2) \sqrt{  (\ell_\nu/r_0)/(n+1) } \right] }\ .
\ee
Thus, in the high-viscosity limit a mode of infinitely long wavelength becomes the most unstable one. The dependence of the dispersion relation on the dimensions and viscosity are depicted in Fig.~\ref{fg:disp}. One can see that the growth rate of the most unstable mode, $\omega(k_{\rm max})$, is also a decreasing function of viscosity, which represents the slowing down of motion by the viscosity. For reference, a derivation of the growth rate for the inviscid case not relying on our formulation in Sec.~\ref{sec:1d} is given in appendix~\ref{sec:inviscid}.

It is interesting to see a long-wavelength limit with the viscosity kept arbitrary.\footnote{The author thanks R.~Emparan for his suggestion to see this limit. A relevant discussion on $k^2$-term can be found in Sec.~3 of Camps \et~\cite{Camps:2010br}, though their approach to the Gregory-Laflamme instability itself is different from ours.} Up to the quadratic order of $kr_0$ ($\ll 1$), growth rate (\ref{eq:disp}) becomes
\be
	\omega
	\simeq
	\omega_0
	\sqrt{ \frac{n}{n+1} }
	(kr_0)
	\left[
		1
		-
		\frac{ n+2 }{ \sqrt{ n(n+1) } }
		\sqrt{ \frac{ \ell_\nu }{ r_0 } } \; ( kr_0 ) 
	\right]\ .
\label{eq:smallk-limit}
\ee
Thus, one can see that the viscosity is responsible for the quadratic term, which is consistent with the expression for the low-viscosity limit in Eq.~(\ref{eq:hilow-limits}). It is noted that expression (\ref{eq:smallk-limit}) does not change if one uses the `incomplete' expression for the mean curvature, Eq.~(\ref{eq:kappa3}).

\section{Discussion}
\label{sec:conc}

We have obtained the set of ($1+1$)-dimensional equations, (\ref{eq:master1}) and (\ref{eq:master2}), for the velocity in the axial direction $v(t,z)$ and height function $h(t,z)$, describing the dynamics of axially symmetric thin flows of non-relativistic and incompressible fluids, where the leading-order effects of the homogeneity in the radial direction are taken into account by series expansion (\ref{eq:expansion}). In spite of the significant simplifications (\ie, the reduction of the numbers of independent variables from three to two, and the `disappearance' of the boundary conditions resulting in the reduction of the numbers of equations from five to two), the set of equations have nice properties: it exhibits the mass conservation (\ref{eq:mass-cons}) and the dissipation of total energy by the viscosity (\ref{eq:energy-diss}) in a plausible manner, provided the complete expression for the  mean curvature (\ref{eq:kappa2}) is used. The growth rate of Rayleigh-Plateau instability (\ref{eq:disp}) was derived, producing the expected/reasonable dependence on the spacetime dimensions $d=n+3$ and shear viscosity $\eta=\rho \nu$.

This work has many directions of application and generalization, a few of which are

\begin{itemize}

\item
Equations (\ref{eq:master1}) and (\ref{eq:master2}) with $n=1$ has been used to describe successfully the drop formation \cite{EggersDupont}. In particular, it has been shown that there is a self-similar solution (precisely speaking, one is before the breakup and another is after the breakup) that plays the role of an attractor in the pinching region~\cite{EggersPRL1993,Eggers1995}. The analysis on higher-dimensional self-similar solutions will be reported elsewhere~\cite{UM}.\footnote{Such a self-similar solution describing the breakup takes the form of $v \sim (t_0 -t)^{-1/2} V(\xi)$ and $ h \sim ( t_0-t ) H (\xi)$ for $d=4$. Here, ($t_0,z_0$) is the time and location of the breakup, and $H$ and $V$ are regular functions of similarity variable $\xi = (z-z_0) / (t_0 - t)^{1/2}$. It would be interesting to see whether the same behavior is found in higher dimensions. See~\cite{Lehner:2010pn} for the observation on the similar behavior near the breakup of black-string apparent horizon.}

\item
One of interesting aspects of both the Gregory-Laflamme and Rayleigh-Plateau instabilities is their dimensional dependence of the phase structures~\cite{Sorkin:2004qq,Kudoh:2005hf} and dynamics~\cite{Miyamoto:2008uf} that appears in non-linear regimes. Above a critical dimension, it is expected that there are non-uniform equilibrium configurations of black string and fluid flows that serve as the end point of the instabilities. Thus, it would be interesting to examine the dimensional dependence of the end point of Rayleigh-Plateau instability with Eqs.~(\ref{eq:master1}) and (\ref{eq:master2}).\footnote{Due to the simplicity of this set of equations, rather `economical' simulations would be possible. The time evolution of the instability can be traced, \eg, with the {\tt NDSolve} command in {\it Mathematica} installed on a laptop computer, at least for simple (\eg, periodic) boundary conditions and away from singularities.}

\item
It would be possible to generalize the formulation in this paper to the relativistic and compressible cases such as one in~\cite{Aharony:2005bm,Lahiri:2007ae} since we would have no additional degrees of freedom essentially once an equation of state is provided by the fluid/gravity correspondence.

\end{itemize}

\subsection*{Acknowledgments}

The author would like to thank R.~Emparan, T.~Harada, and F.~Pretorius for useful discussions and comments. This work is in part supported by the Grant-in-Aid for Scientific Research Fund of the Ministry of Education, Culture, Sports, Science and Technology, Japan [Young Scientists (B) 22740176], and by Research Center for Measurement in Advanced Science in Rikkyo University.

\appendix

\section{Instability of inviscid flows}
\label{sec:inviscid}

Let us consider linear perturbations of an inviscid flow ($\nu=0$) that has a uniform velocity in the $r$-direction. For this purpose, it is convenient to work in a boosted frame in which the uniform velocity vanishes. We assume that the perturbation results from a sinusoidal disturbance of the height function given by
\be
	h(t,z) = r_0 [ 1 + \varepsilon(t) \cos (kz) ]\ ,
\ee
where $| \varepsilon(t) | \ll 1$.
This disturbance of the height function leads to those of the pressure and velocity field, where the disturbed pressure would take form of $ p=p_0 + \delta p (t,z,r)$. Here, $p_0$ is a constant determined by the normal-stress balance as $p_0 = n \alpha/r_0 $. For incompressible fluids the pressure operated by Laplacian vanishes in general, $\Delta p = 0$, and therefore we have
\be
	\Delta \delta p = 0\ .
\label{eq:Deltap}
\ee
Writing the perturbation of pressure as $ \delta p = \delta \bar{p}(t) F_k(r) \cos (kz) $, Eq.~(\ref{eq:Deltap}) leads to
\be
	\frac{ \dd^2 F_k }{ \dd r^2 }
	+
	\frac{n}{r} \frac{\dd F_k}{\dd r}
	-
	k^2 F_k = 0\ .
\label{eq:dFdr}
\ee
This equation is solved by the modified Bessel function of the first kind (\eg, \cite{Arfken}),
\be
	F_k(r) = \frac{ I_{(n-1)/2} ( kr ) }{ r^{(n-1)/2} }\ ,
\label{eq:F(r)} 
\ee
where we have assumed the regularity at the axis, $r=0$.
The combination of (\ref{eq:F(r)}) and the perturbation of normal-stress balance (\ref{eq:sb1}) (i.e.\ $\delta p |_{r=r_0} = \alpha \delta \kappa $) determines the time dependence of the pressure perturbation,
\be
	\delta \bar{p}(t)
	=
	-   
	\frac{ \alpha r_0^{ (n-3)/2 } [ n-(kr_0)^2 ]   }
		 { I_{(n-1)/2} (kr_0) }
	\epsilon(t)\ .
\label{eq:deltap}
\ee
\begin{figure}[t!]
	\begin{center}
	\includegraphics[width=8cm]{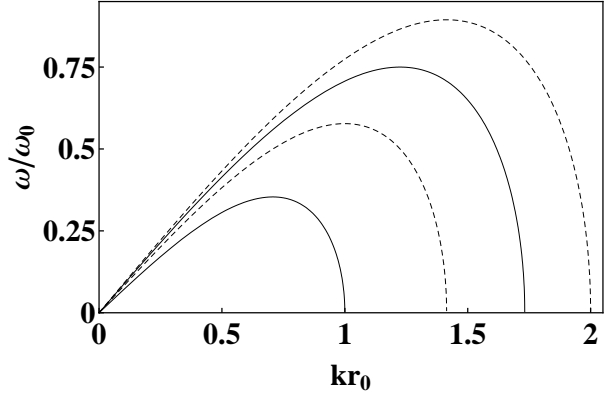}
	\caption{{\sf The growth rate of Rayleigh-Plateau instability for an inviscid flow, in $d=4,5,6,7$ from the bottom to the top.}}
	\label{fg:dispInvis}
	\end{center}
\end{figure}
The perturbation of velocity in the $r$-direction, $\delta v_r(t,z,r)$, at the boundary is immediately obtained from kinetic boundary condition (\ref{eq:kbc}),
\be
	\delta v_r |_{r=r_0}
	=
	\pd_t \delta h
	=
	r_0 \dot{\epsilon}(t) \cos ( kz )\ .
\label{eq:deltavr}
\ee 
The Navier-Stokes equation in the $r$-direction, Eq.~(\ref{eq:ns2}), relates the pressure- and velocity-perturbations,
\be
	\pd_t \delta v_r \Big|_{r=r_0}
	=
	- \frac{ \pd_r \delta p }{ \rho } \Big|_{r=r_0}\ .
\label{eq:Pert-NSr}
\ee
Plugging equations (\ref{eq:F(r)}), (\ref{eq:deltap}), and (\ref{eq:deltavr}) into equation (\ref{eq:Pert-NSr}), we obtain an equation for $\varepsilon(t)$,
\be
	\ddot{\varepsilon}(t)
	=
	\varepsilon(t)
	\frac{ \alpha r_0^{(n-5)/2} [n-(kr_0)^2] }{ \rho I_{(n-1)/2} (kr_0) }
	\pd_r \left[ \frac{ I_{(n-1)/2}(kr) }{ r^{(n-1)/2} } \right]
	\Bigg|_{r=r_0}\ .
\ee
After some calculations with assuming $ \varepsilon(t) \propto e^{\omega t} $, we obtain a dispersion relation of the linear perturbation\footnote{We use the following formula to eliminate the derivative of modified Bessel function,
\be
	\left(
		\frac{1}{z} \frac{ d }{ dz }
	\right)^{m} \left[ z^{-\beta} I_\beta(z) \right]
	=
	(-1)^m z^{-\beta-m} I_{\beta+m} (z)\ ,
\ee
with $m=1$ and $\beta=(n-1)/2$.
}
\be
	\omega^2
	=
	\omega_0^2  
	\frac{ kr_0[ n-(kr_0)^2 ]  I_{(n+1)/2} (kr_0) }
		 { I_{(n-1)/2}(kr_0) }\ ,
\label{eq:disp-inv}
\ee
which reproduces the result in~\cite{Cardoso:2006sj,Caldarelli:2008mv}.
The dispersion relation for the inviscid case, Eq.~(\ref{eq:disp-inv}), in several dimensions are depicted in Fig.~\ref{fg:dispInvis}.



\end{document}